\documentclass[aps,twocolumn,prb,floatfix,showpacs,superscriptaddress]{revtex4-1}
\usepackage{amsfonts}
\usepackage{amsmath, amssymb}    
\usepackage{graphicx}   
\usepackage{natbib,hyperref}
\usepackage{textcomp}
\usepackage{verbatim}   
\usepackage{color}      
\usepackage{subfigure}  
\usepackage{hyperref}   
\usepackage{wrapfig}
\usepackage{xcolor}
\usepackage{soul}
\usepackage{siunitx}
\usepackage{soul}
\usepackage{braket}
\usepackage{textgreek}
\usepackage{textcomp}
\usepackage[version=4]{mhchem}

\usepackage{epstopdf}

\definecolor{cream}{RGB}{222,217,201}

\begin{document}

	\title{Dielectric function of \ce{CuBr_xI_{1-x}} alloy thin films}
	\author{Michael Seifert}
	\altaffiliation{These authors contributed equally}
	\affiliation{Institut f\"{u}r Festk\"{o}rpertheorie und Optik, Friedrich-Schiller-Universit\"{a}t Jena, Max-Wien-Platz 1, 07743 Jena, Germany }
	\affiliation{European Theoretical Spectroscopy Facility}
	\author{Evgeny Kr\"{u}ger}
	\altaffiliation{These authors contributed equally}
	\affiliation{Felix-Bloch-Institut f\"{u}r Festk\"{o}rperphysik, Universit\"{a}t Leipzig, Linn\'{e}str. 5, 04103 Leipzig, Germany }
	\author{Michael S. Bar}
	\affiliation{Felix-Bloch-Institut f\"{u}r Festk\"{o}rperphysik, Universit\"{a}t Leipzig, Linn\'{e}str. 5, 04103 Leipzig, Germany }
	\author{Stefan Merker}
	\affiliation{Institut f\"{u}r Anorganische Chemie, Universit\"{a}t Leipzig, Johannisallee 29, 04103 Leipzig, Germany }
	\author{Holger von Wenckstern}
	\affiliation{Felix-Bloch-Institut f\"{u}r Festk\"{o}rperphysik, Universit\"{a}t Leipzig, Linn\'{e}str. 5, 04103 Leipzig, Germany }
	\author{Harald Krautscheid}
	\affiliation{Institut f\"{u}r Anorganische Chemie, Universit\"{a}t Leipzig, Johannisallee 29, 04103 Leipzig, Germany }
	\author{Marius Grundmann}
	\author{Chris Sturm}
	\affiliation{Felix-Bloch-Institut f\"{u}r Festk\"{o}rperphysik, Universit\"{a}t Leipzig, Linn\'{e}str. 5, 04103 Leipzig, Germany }
	\author{Silvana Botti}
	\affiliation{Institut f\"{u}r Festk\"{o}rpertheorie und Optik, Friedrich-Schiller-Universit\"{a}t Jena, Max-Wien-Platz 1, 07743 Jena, Germany }
		\affiliation{European Theoretical Spectroscopy Facility}

	\begin{abstract}	
		We study the dielectric function of \ce{CuBr_xI_{1-x}} thin film alloys using spectroscopic ellipsometry in the spectral range between \SI{0.7}{eV} to \SI{6.4}{eV}, in combination with first-principles calculations based on density functional theory. Through the comparison of theory and experiment, we attribute features in the dielectric function to electronic transitions at specific k-points in the Brillouin zone. The observed bandgap bowing as a function of alloy composition is discussed in terms of different physical and chemical contributions. The  band splitting at the top of the valence band due to spin-orbit coupling is found to decrease with increasing Br-concentration, from a value of \SI{660}{meV} for CuI to \SI{150}{meV} for CuBr. This result can be understood considering the contribution of copper d-orbitals to the valence band maximum as a function of the alloy composition.
	\end{abstract}

	\maketitle
	\section{introduction}
		Transparent conductive materials (TCMs) succeed in bringing together high optical transparency in the visible part of the electromagnetic spectrum and good electric conductivity. They are therefore essential building blocks for future innovative applications in the field of transparent optoelectronics \cite{MoralesMasis.2017,Liu.2018,Kateb_2016}, such as transparent thin film transistors~\cite{Liu.2018}, transparent electrodes~\cite{Granqvist_1993}, electrochromic displays~\cite{Kateb_2016} and solar windows~\cite{Yang_2017}. While n-type metal-oxide semiconductors, like, e.g., amorphous ZTO~\cite{	lahr2019full,lahr2020ultrahigh,lahr2020all}, InGaZnO\cite{Nomura.2003,Nomura.2004,Thomas.2013}, indium tin oxide (ITO)~\cite{Sakamoto_2018,Si.2020} are widely used for thin film transistors and as electrodes in transparent commercial optoelectronic devices, achieving high-performance transparent p-type materials, that can be deposited as thin films, is still an open challenge~\cite{AoLiu_2021, Hu_2020, AoLiu_2020}. 
		
		Promising p-type TCMs, such as CuAlO\textsubscript{2}~\cite{Kawazoe.1997}, SnO\textsubscript{4}~\cite{Ogo.2008}, NiO~\cite{Liu.2019,Gagaoudakis.2020}, and copper halides (CuI, CuBr, CuCl)~\cite{Grundmann_2013,Yang_2017,Zhu.2019,Storm.2020, Chang.2021} have attracted increasing interest in the last years. In particular, copper halides have more delocalized holes at the valence band maximum (VBM) compared to oxides, and therefore higher mobilities. This is a consequence of the fact that the electronegativities of Cl, Br and I atoms are smaller than the one of oxygen~\cite{Yamada_2016, Hu_2020}. The zincblende phase of CuI, $\gamma$-CuI, has particularly interesting electronic properties and therefore has been attracting increasing interest~\cite{Grundmann_2013,Annadi.2019,Storm.2020,AoLiu_2021}. With a bandgap of around $3.1~\mathrm{eV}$~\cite{Grundmann_2013, Krueger.2021}, large exciton binding energies of $62~\mathrm{meV}$~\cite{Krueger.2021} and high hole mobilites~\cite{Chen.2010} of $\mu >$ \SI{40}{cm^2 V^{-1}s^{-1}} in bulk single crystals, $\gamma$-CuI is a suitable transparent p-type material for optoelectronic applications. Until now, the application potential of CuI has already been demonstrated in transparent p-n heterojunctions~\cite{Yang_2016}, thin film transistors~\cite{Choi_2016,Liu.2018,Tixier_2016}, light-emitting diods~\cite{Ahn_2016, Baek_2020}, perovskite solar cells~\cite{Christians_2014,Yu_2018, Matondo_2021}, UV-photodetectors~\cite{Yamada.2019} and thermoelectric devices~\cite{Yang_2017}. Additionally, compatibility with several $n$-type materials has been proven by building prototype devices \cite{Yang_2016, Schein_2013, Ding_2012, Lee_2021}. 

		However, excessive hole densities of $p\approx\left( 10^{18} - 10^{19} \right)~\mathrm{cm^{-3}}$ as reported in Ref.~\onlinecite{Storm.2020} typically observed in CuI thin films, are disadvantageous for active device applications~\cite{AoLiu_2020,Yamada.2020}. Exploring ways to control hole densities of CuI-based thin films is therefore crucial for technological progress. An improved p-type conductivity has been recently demonstrated using thermal treatment~\cite{Yamada.2016,Liu.2018} or by Zn-doping~\cite{Liu.2020}. An alternative promising optimization route based on \ce{CuBr_xI_{1-x}} alloys was reported by Yamada et al.~\cite{Yamada.2020}: the hole density of the alloy was shown to be tunable over several orders of magnitude, thanks to the fact that the (0/-) charge transition level of the Cu vacancy in CuBr is deeper in comparison to the one of CuI\cite{Yamada.2020}. The effective hole mass of CuBr is considered to be very similar to the one of CuI, and therefore the mobility of positive charge carriers in \ce{CuBr_xI_{1-x}} alloys can  be reduced by the alloy disorder~\cite{Yu.1973,Bouhafs.1998}.  Also \ce{CuBr_xI_{1-x}} thin film were successfully applied to build transparent p-n-junctions ~\cite{Mori.2019}, solar cells~\cite{Rajani.2013,Bhargav.2019} and thin film transistors~\cite{Zhu.2019}. In contrast of the binary compounds, CuI~\cite{Blacha.1986,Song.1967,Seifert_2021,Krueger_2018,Krueger.2021} and  CuBr~\cite{Goldmann.1977,Suga.1976,Gao.2018}, only few theoretical and experimental results are reported in literature for \ce{CuBr_xI_{1-x}}  alloys~\cite{Cardona_1963,Bouhafs.1998,Yamada.2020,Tanaka.2001,Chang.2021}. Although absorption and reflectivity spectra of \ce{CuBr_xI_{1-x}} were already published in the 60s~\cite{Cardona_1963}, a dedicated investigation of the electronic structure and optical properties as a function of alloy composition is still lacking. 
		
		As a comprehensive understanding of the optical response of \ce{CuBr_xI_{1-x}} alloys and the underlying electronic transitions is crucial for the design and fabrication of transparent optoelectronic devices, we investigate in this work, combining experimental and computational methods, the dependence on alloy composition of the electronic band structure and dielectric function of zincblende \ce{CuBr_xI_{1-x}}. 
		
The manuscript is structured as the following: In the main manuscript the main physical effects are discussed. A discussion and approximation of effects which are important to compare the experimental results with the computational like for example the effect of temperature and the excitonic binding energy is provided in the supplementary material.

	\section{Experimental Methods}
		The investigated \ce{CuBr_xI_{1-x}} thin films were deposited on amorphous quartz glass substrates at room temperature by combinatorial pulsed laser deposition (PLD) using a powder-based elliptically segmented target~\cite{Wenckstern.2020}. For CuI, we used commercial powder (CuI: 98\%, Carl Roth), while CuBr powder was crystallized at room temperature from the hot supernatant solution of copper(I) bromide (98\%, Acros Organics) in hydrobromic acid (48\%, extra pure, Acros Organics) with Cu (99,9\% Alpha Aeser) for the reduction of Cu$^{2+}$-ions. Handling of CuBr was done under \ce{N2}-atmosphere using standard Schlenk techniques and a glovebox to prevent oxidation. A detailed description of the PLD setup as well as recently reported PLD deposition of CuI thin films can be found in Ref.~\onlinecite{Storm.2020}. In addition, the corresponding uniform powder targets were used for the deposition of the binary CuI and CuBr films, which served as references. Details on deposition as well structural and electrical properties of alloyed \ce{CuBr_xI_{1-x}} thin films will be published elsewhere~\cite{Bar.2022}. The film thicknesses of the  \ce{CuBr_xI_{1-x}} thin films investigated in our work are around \SI{1}{\micro m}, independently of the alloy compositions. Since Cu-halides are known to suffer from oxidation when exposed to the atmosphere, an additional \ce{Al_2O_3} capping layer \mbox{(d $\approx$ \SI{140}{nm})} was deposited in situ on top of the \ce{CuBr_xI_{1-x}} layer using a sintered \ce{Al_2O_3} target, as reported recently for CuI~\cite{Storm.2021}. The \ce{N2} partial pressure and growth temperature used for the deposition of \ce{Al_2O_3} thin films were chosen to be \mbox{p(\ce{N2}) = \SI{3e-3}{mbar}} and \mbox{T = \SI{300}{K}}, respectively, i.e. identical to those used for the deposition \ce{CuBr_xI_{1-x}} layers.

		The crytallographic structure of the deposited \ce{CuBr_xI_{1-x}} thin films was analyzed by X-ray diffraction (XRD) measurements in Bragg-Brentano geometry using Cu \ce{K_\alpha} radiation. The observed peaks in the 2$\Theta-\omega$ scans (see Fig. S1 (a) in the supplementary material) can be attributed exclusively to the zincblende structure, suggesting solid-solution-like, single-phase thin films. While in the I-rich alloy compositions up to \mbox{x $\approx$ 0.7} the films are preferentially oriented along the (111) direction, Br-rich films exhibit polycrystalline structure. The lattice constant determined from the XRD scans (see Fig. S1 in the supplementary material) was used to estimate the alloy composition of the \ce{CuBr_xI_{1-x}} alloy using Vegard's law~\cite{Vegard.1921} .
	
		\begin{figure*}[!htbp]
			\centering
			\includegraphics[width=\textwidth]{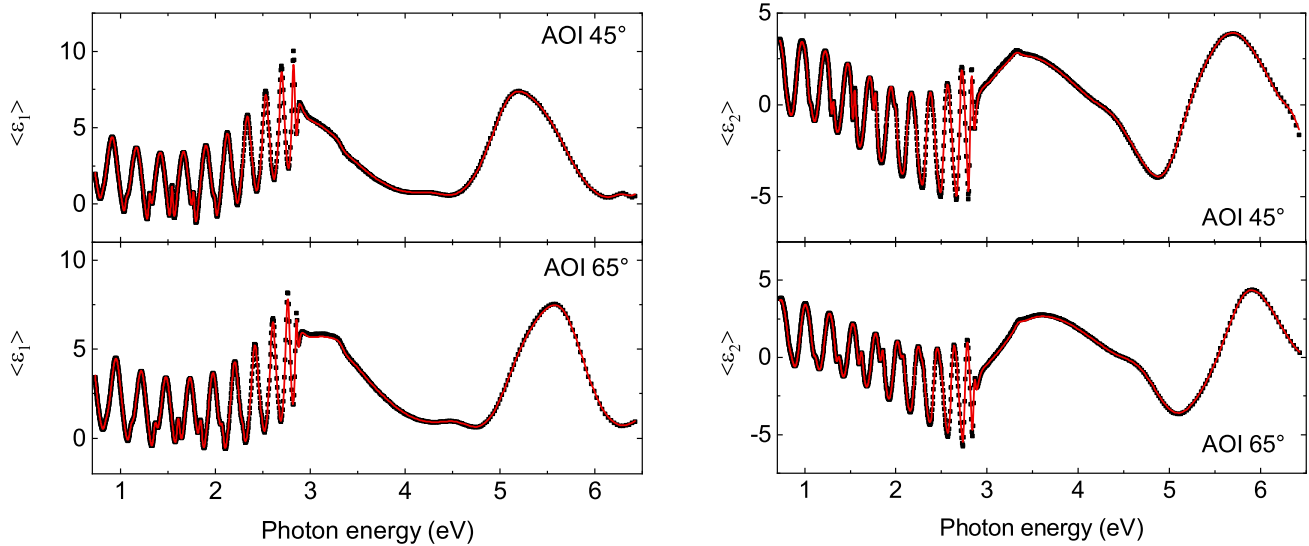}
			\caption{Experimental (black squares) and calculated (solid red lines) spectra of the pseudo dielectric function $<\varepsilon_1>$ (left column) and $<\varepsilon_2>$ (right column) by means of the transfer matrix technique shown exemplary for angles of incidence (AOI) of \SI{45}{\degree} and \SI{65}{\degree} of \ce{CuBr_{0.5}I_{0.5}} thin film.} 
			\label{fig:ellipsofitexample}
		\end{figure*}

		The dielectric function in the spectral range of \SIrange{0.7}{6.4}{eV} was determined by means of standard spectroscopic ellipsometry using a dual rotating compensator ellipsometer (RC2, J.A. Woollam). The ellipsometry data were analyzed by means of the transfer matrix technique for a layer stack model, consisting of a semi-infinite substrate, two layers for the \ce{CuBr_xI_{1-x}} thin film and the \ce{Al2O3} capping film, and a layer describing the surface roughness, bound by air as the ambient. For the substrate, we used previously determined optical constants of amorphous \ce{SiO2}. The dielectric function of the \ce{CuBr_xI_{1-x}} film was modeled numerically using a Kramers-Kronig consistent B-spline approach~\cite{Johs.2008}. The depolarization observed in the measured data could be described by an inhomogeneity of the film thickness of the \ce{CuBr_xI_{1-x}} film, which is between 1\% and 2\%. The surface layer was modelled using the Bruggeman effective medium approach~\cite{Bruggeman.1935}, mixing  the dielectric function of the \ce{Al2O3} thin film with void with ratio of 1:1. Fig.\ref{fig:ellipsofitexample} shows exemplary experimental spectra of the \ce{CuBr_xI_{1-x}} thin film with $x=0.5$ along with the numerical model approximation, demonstrating good agreement with the measured spectra.

	\section{Theoretical and computational methods}	
		All calculations were performed in the framework of density functional theory (DFT). We used the Vienna {\it ab initio} simulation package VASP~\cite{Kresse_1996, kresse_joubert_1999} that implements the projector augmented wave method~\cite{Bloechl_1994}. The $4s$, $3p$ and $3d$ electrons of Cu, as well as the $s$ and $p$ electrons of the outermost shells of I and Br were explicitly treated as valence electrons. A plane-wave basis set with a cutoff energy of 640~eV and a $\mathbf{k}$-point grid of 8$\times$8$\times$8 for the binary compounds, and 4$\times$4$\times$4 for the ordered alloys were used, to assure an error in the total energy of less than \SI{1}{meV/atom}. To focus on the effects of chemical substitution rather than on configurational disorder and to limit the computational costs, we decided to simulate ordered alloys with 2$\times$2$\times$2 supercells (i.e. 8 Cu atoms and 8 halogen atoms). More information on the alloys structures is given in Sec.~I and II of the supplementary material. We considered all possible non-equivalent ways to arrange Br and I atoms at a fixed composition and we optimized the corresponding crystal geometries. The relaxation was done using the PBEsol exchange-correlation functional~\cite{Perdew_2008}, a revised Perdew-Burke-Ernzerhof generalized gradient approximation (GGA) for solids. All forces were relaxed until they were smaller than \SI{1}{meV/ \angstrom}. The cell with the lowest energy at a certain concentration was then determined. This cell was then used for all further calculations at this concentration. We remark that the energetic distance to the other configurations is small enough that we expect them to play a role at room temperature (which was the deposition temperature) so that a disordered alloy should be formed. However, the electronic structure is overall similar for all those non-equivalent configurations, as we show in Sec.~VI of the supplementary material. We can therefore use the results of the lowest energy configurations to discuss general trends of the band structure and effective masses to compare with the experiments. When considering a single configuration is not enough to interpret the trend of experimental observations, we will resort to simple ensemble averages of the configurations with equal stoichiometry. A proper treatment of effects of disorder would require expensive calculations we would leave to later studies which are beyond the scope of the present work. Since Kohn-Sham band structures calculated with the PBEsol functional severely underestimate the bandgap, the hybrid functional PBE0~\cite{PBE0_1996} was used with the inclusion of spin-orbit coupling (SOC) for a correct description of the bandgap, as reported previously for CuI~\cite{Seifert_2021}. For the calculation of the dielectric function we applied the Fermi golden rule in the independent-particle approximation. A Lorentzian broadening of \SI{0.1}{eV} was applied. For the calculation of the spectra the $\mathbf{k}$-point grid was increased to 16$\times$16$\times$16. By extracting the $\mathbf{k}$-dependent transition matrix elements we were able to calculate the contribution to the dielectric function coming from transitions close to specific high-symmetry points in the Brillouin zone. To define a limited region around a certain $\mathbf{k}$-point, we chose a cube with a length of 6/16 (considering the sampling of the Brillouin zone by the chosen grid of 16$\times$16$\times$16) in units of reciprocal lattice vectors and put the corresponding $\mathbf{k}$-point in the middle of the cube. All $\mathbf{k}$-points inside this cube were defined to be in an environment around this specific $\mathbf{k}$-point. We also considered the contribution to the dielectric function of a limited number of bands. 
		
		We calculate the optical transition matrix elements $M_{cv\textbf{k}}$ as~\cite{Matthes_2013}
		\begin{align}
			M_{cv\textbf{k}} &= \lim_{\textbf{q}\rightarrow 0} \frac{e}{|\textbf{q}|} \braket{c\textbf{k} | e^{i\textbf{q}\cdot \textbf{r}} | v(\textbf{k}+\textbf{q})}
		\end{align}
		and link them to the corresponding matrix element $\left\langle c \textbf{k} \middle| \frac{\textbf{q}}{|\textbf{q} |} \textbf{p} \middle|  v\textbf{k} \right\rangle$ using the dipole approximation:
		\begin{align}
			\left\langle c \textbf{k} \middle| \frac{\textbf{q}}{|\textbf{q} |} \textbf{p} \middle|  v\textbf{k} \right\rangle &= \frac{m}{\hbar e} \left[  \epsilon_c(\textbf{k}) - \epsilon_v(\textbf{k})  \right] M_{cv\textbf{k}}
		\end{align}

Excitonic effects would induce a mixing of the independent-particle transition matrix elements, leading to shifts of peak postions and modifications of the oscillator strength. However, in these p-type materials we expect screening of the electron-hole interaction due to free charges at the top of the valence band, and we can reasonably expect overall small excitonic corrections. Also considering the very high computational cost, we decided to neglect these corrections in our calculations and to evaluate the excitonic binding energy for the lowest transition at the $\Gamma$-point with an effetive mass model described in Sec.~V of the supplementary information.

	\section{Results and discussion}
	\subsection{Dielectric function of $\text{Cu}\text{Br}_\text{x}\text{I}_{\text{1-x}}$}
		\begin{figure*}[!htbp]
			\centering
			\includegraphics[width=\textwidth]{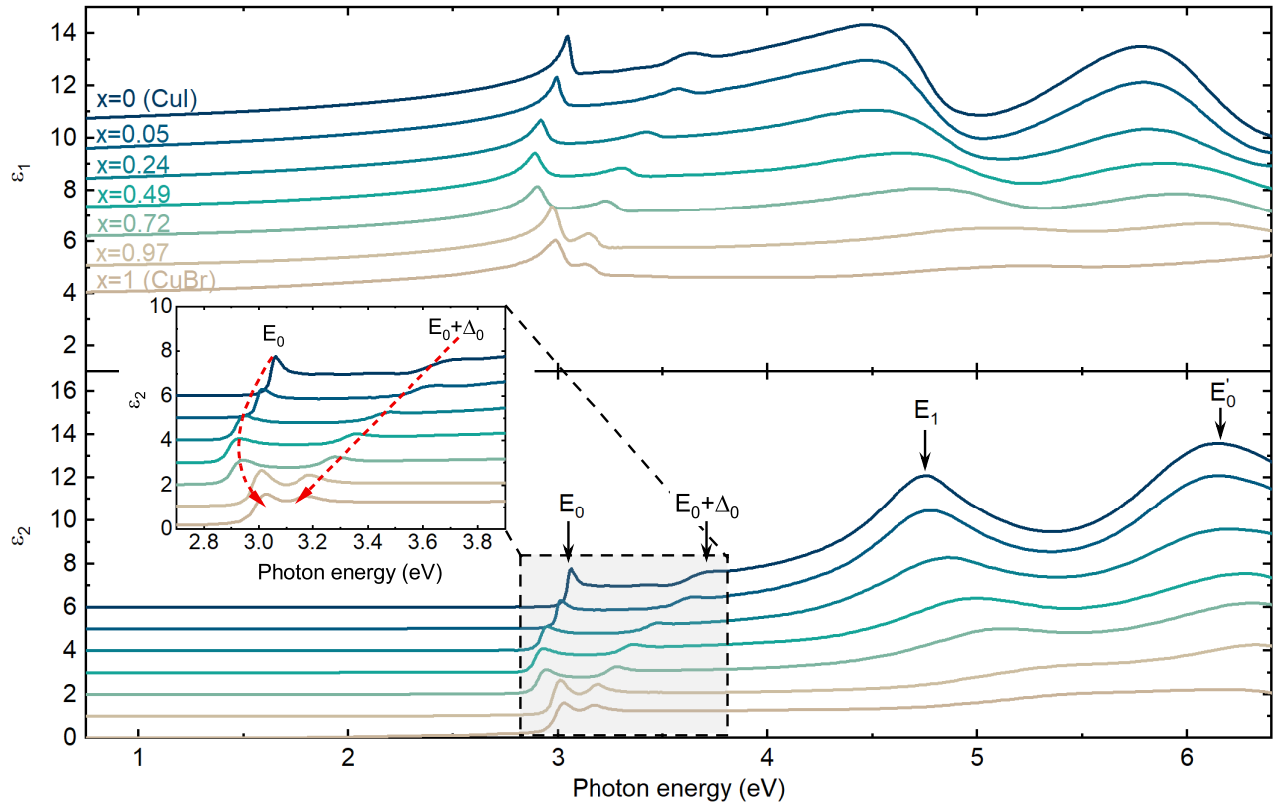}
			\caption{Experimental spectra of the real ($\varepsilon_1$) and imaginary ($\varepsilon_2$) part of the dielectric function of \ce{CuBr_xI_{1-x}} thin films as a function of the alloy composition for $0 \leq x \leq 1$. The main peaks are marked with vertical arrows and labeled with E\textsubscript{0}, E\textsubscript{0}+$\Delta$\textsubscript{0}, E\textsubscript{1} and $\text{E}'_\text{0}$. The spectra have been offset vertically for the sake of clarity. The inset shows a zoom of $\varepsilon_2$ in the vicinity of the excitonic resonances at the \textGamma-point. The dashed red arrows highlight the bowing of the excitonic resonance E\textsubscript{0} and the monotonic shift of the E\textsubscript{0}+$\Delta$\textsubscript{0} transition, respectively. }
			\label{fig1}
		\end{figure*}

		\begin{figure*}[!htbp]
			\includegraphics[width=\textwidth]{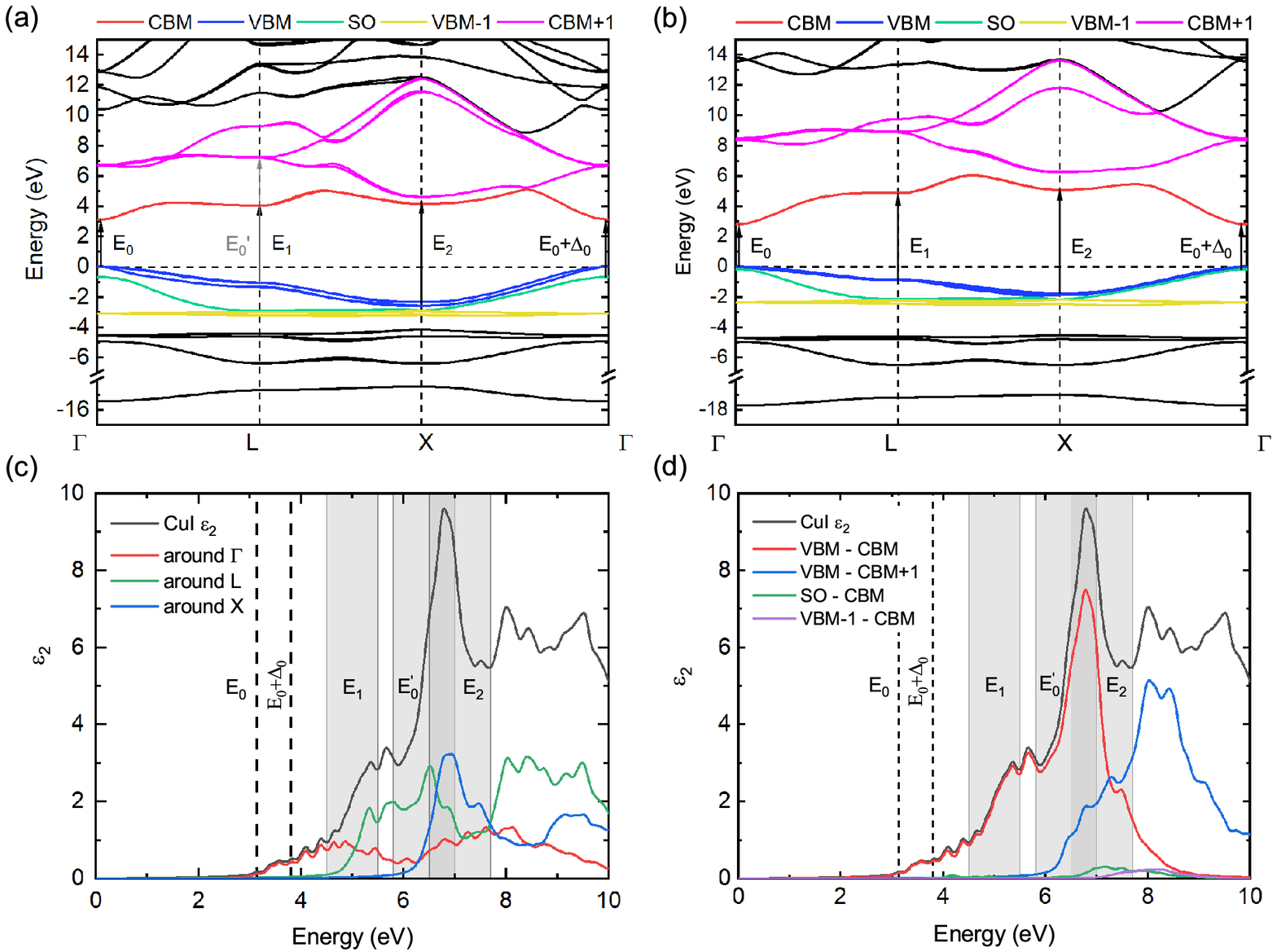}
			\caption{Top: band structures of CuI (a) and CuBr (b) calculated using PBE0 including of spin-orbit coupling (SOC). The main electronic transitions around symmetry point in the Brillouin zone are indicated by vertical black arrows and labeled consistently with the notation of Fig.\ref{fig1}. The colors distinguish groups of bands. Bottom: calculated absorption spectra (i.e., imaginary part of the dielectric function) of CuI. The contributions of all bands at selected symmetry points, as well as contributions of transitions involving different groups of bands over the entire Brillouin zone are shown in (c) and (d), respectively. The definition of how the region around the symmetry points was set is given in the computational details. Analogous calculated spectra for CuBr can be found in Sec. III of the supplementary material.}
			\label{fig2}
		\end{figure*}

		The effect of the alloy composition on the dielectric function of the \ce{CuBr_xI_{1-x}} alloy for $0 \leq x \leq 1$ is presented in Fig.~\ref{fig1}. As expected, the thin films are almost transparent in the visible spectral range. The first transition E$_0$ occuring at \SI{3.06}{eV} for CuI and \SI{3.03}{eV} for CuBr can be attributed to excitonic excitations at the fundamental bandgap at the \textGamma-point. In the case of CuI, the calculated quasi-particle gap of \SI{3.07}{eV} (see Fig.~\ref{fig2} (a)) matches the experimental value under consideration of the exciton binding energy of \SI{62}{meV} for CuI ~\cite{Krueger.2021} excellently well. For CuBr, the calculated quasi-particle gap is approximately \SI{2.7}{eV} (see Fig.~\ref{fig2} (b)) and is therefore slightly lower in comparison to the experimental value of \SI{3.03}{eV}, considering the literature value for the excitonic binding energy of \SI{108}{meV}\cite{Ueta.2012}. Still, the here obtained values are in a good agreement with the experiment and the difference is smaller than what can be expected by DFT\cite{Borlido.2019}.

		We note that the bandgap of these binary compounds are very similar and are in both cases significantly smaller than from the extrapolation of the corresponding transitions of the isoelectronic sequence~\cite{Herman.1955} (GaAs and ZnSe for CuBr and GaSb and ZnTe for CuI). Such behavior was already suggested to be caused by the strong impact of the Cu 3d-orbitals\cite{Cardona_1963} as it is confirmed by recent first-principles calculations ~\cite{Yu.2022}. The effect of the p-d-hybridization of the valence bands of the \ce{CuBr_xI_{1-x}} alloy will be discussed in section \ref{sec:SOC} in more detail. The origin of the non-linear behavior of the E\textsubscript{0} transition energy is treated in section ~\ref{sec:bandgap}.

		The resonance labeled with E$_0+\Delta_0$ in Fig.~\ref{fig1} is also attributed to transitions at the \textGamma-point but involving the split-off component of the top valence band and the lowest conduction band.  From our DFT-calculations we obtain spin-orbit splittings of \SI{670}{meV} and \SI{170}{meV} for CuI and CuBr, respectively, in good agreement with previous experimental results for the binary compounds ~\cite{Cardona_1963, Tanaka.2001}. 
		
		The increasing underestimation of the calculated spin-orbit splitting $\Delta_0$ for Br-rich alloy compositions compared to the experimental values can be explained by excitonic effects for the $E_0$ and $E_0+\Delta_0$ transitions. A more detailed consideration of the spin-orbit splitting as a function of the alloy composition is presented in section \ref{sec:SOC}. Finally, we would like to emphasize that both transitions $E_0$ and $E_0+\Delta_0$ are expected to be of excitonic nature, which is also confirmed by the observed line shape of the dielectric function in the vicinity of the corresponding resonances.

		The next higher-energetic transition occurring between \SI{4.5}{eV} and \SI{5.5}{eV}, labeled with E\textsubscript{1}, was attributed in our earlier study on CuI to transitions between the top VB and the lowest CB around the L-point~\cite{Krueger_2018}. This is verified by our present calculations (see Fig.~\ref{fig2}) (c). (The corresponding pictures for CuBr to Fig.~\ref{fig2} (c) and (d) are provided in Fig.~S3. of the supplementary material.)  The corresponding peak in the calculated $\varepsilon_2$ spectra appears ca. \SI{0.2}{eV} above the energy spacing between the top VB and the lowest CB of ca. \SI{5.1}{eV} directly at the L-point. From this we conclude that the E\textsubscript{1} peak is due to transitions close to the L-point and not only by transitions directly at the Brillouin zone boundary, confirming the suggestion made in the past for other zincblende-like semiconductors~\cite{Brust.1962}. The peak position obtained from the experimental data is about \SI{4.8}{eV} and agrees reasonably with calculated band structure. Although, the exact spectral peak position of the E\textsubscript{1} transition is hard to extract for Br-rich alloy compositions due to large broadening, we observe experimentally a shift to higher energies with increasing Br-incorporation (see Fig.~\ref{fig1}) ending up at ca. \SI{5.4}{eV} for binary CuBr. The increasing transition energy with increasing Br-content is also confirmed by our calculations. From the band structure of CuBr shown in Fig.~\ref{fig2} (b) we see that the transition directly at the L-point would be at 5.7~eV. The main peak in the calculated spectrum (see Fig. S3 in supplementary material) is located at \SI{5.9}{eV}. So again, here the calculations indicate a slightly larger peak position. We note that although the transition at the L-point involving the split-off component of the top VB (E\textsubscript{1}+$\Delta_1$) cannot be resolved in our work due to significant overlap with the main E\textsubscript{1} peak at room temperature, the calculated spin-orbit splitting exactly at the L-point $\Delta_1$ is \SI{304}{meV} for CuI and \SI{4}{meV} for CuBr. In a small environment around L (the same definition as introduced in the computational method, but now with a length for the cube of 2/16 with the L-point in the center), these values rise up to \SI{336}{meV} for CuI and \SI{124}{meV} for CuBr. This is in reasonable agreement with the estimation that this value should be around 2/3 of $\Delta_0$ \cite{Cardona_1963}.

		At the high energy side of the measured dielectric function spectra a further transition can be observed and is labeled with $\text{E}'_\text{0}$. We note that at low temperatures a triplet peak structure was observed for CuI in this spectral region~\cite{Krueger_2018} However, due to the significant spectral overlap of the peaks due to thermal broadening we cannot distinguish between the different transitions in this work. Based on the isoelectronic sequence, the corresponding transitions were assigned lately to transitions at the \textGamma-point~\cite{Cardona_1963}. However, based on a similar temperature dependence of this peak and the E\textsubscript{1} peak, it has recently been suggested, that it could also originate from transitions between the split-off component of the top VB and the second CB in the vicinity of the L-point~\cite{Krueger_2018}, similarly to the interpretation of Song et al.~\cite{Song.1967}. In our calculations for both CuI and CuBr  we also observe contributions from the vicinity of the L-point due to transitions from VBM to CBM+1 (the group of bands at the $\Gamma$-point above the CBM) at energies below the next high-energy transition E$_2$ close to the X-point (compare Fig~\ref{fig2}). Although, these two peaks $\text{E}'_\text{0}$ and E$_2$ strongly overlap, it is still possible to see that the contribution of transitions around the \textGamma-point at these energies are smaller. Taking into account the previously reported temperature-dependent dielectric function of CuI~\cite{Krueger_2018}, we thus conclude that the $\text{E}'_\text{0}$ transitions indeed occur mainly near the L-point.

		Furthermore, we examine the structure of the low energy valence bands in more detail. While  monoatomic crystals in the diamond structure with covalent bonding, such as Si or Ge, do not have a gap between the lowest VB at the X-point, in biatomic zincblende semiconductors with partially ionic bonding, an energy gap is observed at this point, which is sometimes called an asymmetric gap~\cite{Chelikowsky.1989}. For CuI, we obtain a value of \SI{8.1}{eV} for this asymmetric gap in good agreement with previous band structure calculations~\cite{Wang.2011}. In the case of CuBr, due to the higher ionicity of CuBr (f=0.735~\cite{Phillips.1970}) compared to CuI (f=0.692~\cite{Phillips.1970}), we obtain, as expected, a larger value of ca. \SI{10.6}{eV}. Therefore, we note that our results are both comparable to the semi-empirical tight-binding calculations of Bouhafs et al.~\cite{Bouhafs.1998} and fit the general trend observed for other zincblende-like compound semiconductors with increasing ionic bonding character. 

		Regarding the real part of the dielectric function, the calculated values of CuI and CuBr at \SI{0.8}{eV} are 4.36 and 3.66, respectively, which is in good agreement of 4.77 and 4.07 of the experimental data. 

		In conclusion the simulated spectra are in good agreement with the experimental results and give insight on the origin of the spectral features.  
In the following we focus the discussion on the dependence of the bandgap energy and the spin-orbit splitting as a function of the alloy compositon.

	\subsection{Bandgap bowing}
	\label{sec:bandgap}
		\begin{figure}[!htbp]
			\centering
			\includegraphics[width=0.5\textwidth]{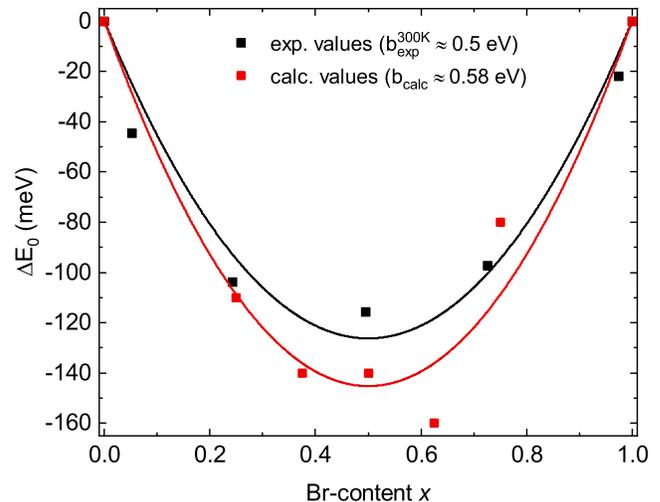}
			\caption{Energy difference $\Delta E_\text{0}(x) = E_\text{0}(x) - E_\text{lin}(x)$ between the measured transition energy E\textsubscript{0} and the weighted average of the ternary transition energies $E_\text{lin}$ as a function of alloy composition. The black symbols represent the experimental values at \SI{300}{K}, while the red symbols represent the values resulting from the calculated band structures for the corresponding alloy compositions for the lowest energy configurations. The solid lines, represent the model fit using $bx(1-x)$.}
			\label{fig3}
		\end{figure}

		Apart from the slight difference in the bandgaps of the two binary compounds, it is evident from Fig. \ref{fig1} that the evolution of the E\textsubscript{0} transition energy does not exhibit a monotonic dependence on the alloy composition. Following the approach presented by Cardona~\cite{Cardona_1963}, we decompose the energetic position of E\textsubscript{0} excitonic peak as a function of alloy composition into a linear and a quadratic term:
		\begin{equation}
			\mathrm{E}_0^{\ce{CuBr_{x}I_{1-x}}}=\left[x\mathrm{E_0}^{\ce{CuBr}}+(1-x)\mathrm{E_0}^{\ce{CuI}}\right]-\left[bx(1-x)\right],
		\end{equation} 

		where E\textsubscript{0}\textsuperscript{\ce{CuI}} and  E\textsubscript{0}\textsuperscript{\ce{CuBr}} are the optical transition energies for the binary compounds, $x$ denotes the Br-content and $b$ is the bowing parameter. Fig.~\ref{fig3} shows the latter nonlinear contribution of the E\textsubscript{0} transition energy. This results in a bowing parameter of about \mbox{$b\approx\SI{0.50}{eV}$} for the E\textsubscript{0} transition, which is comparable with the results known from the literature \cite{Cardona_1963,Bouhafs.1998}. 

The red symbols in Fig.~\ref{fig3} show the DFT results of the lowest energy configuration. As we discuss later in detail included are there structural contributions (both the cell volume and the internal positions were relaxed). Neglected are excitonic contributions as well as disordered contributions. The bowing parameter for this setting is \mbox{$b\approx\SI{0.58}{eV}$}. However, when considering the other ordered configurations the bowing becomes larger. The result of the ensemble average is \mbox{$b\approx\SI{1.00}{eV}$}. A full discussion of the influence of the ordered configurations together with other effects one needs to consider is given in the supplementary material. Furthermore, we note that due to the high exciton binding energy of \SI{62}{meV} for CuI\cite{Krueger.2021} and \SI{108}{meV} for CuBr~\cite{Ueta.2012}, the optical transitions exhibit a strong excitonic character even at room temperature. Thus, since the exciton binding energy is expected to depend on the alloy composition, it is not immediately obvious that the theoretically computed bowing of the bandgap must match the experimentally observed bowing of the E\textsubscript{0} transition. Indeed, we find a slightly non-linear increase in exciton binding energy with increasing Br-incorporation which were calculated within a hydrogenic model. (see Sec.V in supplementary material). Considering this non-linearity the bandgap bowing parameter can be estimated to a value of \SI{0.43}{eV}, being thus approximately 11\% smaller compared to the bowing behavior of the excitonic $E_0$ transition shown in Fig. \ref{fig3}. In addition to the excitonic effects, the temperature dependence of the bandgap for different alloy compositions must also be considered, especially because it is known that the temperature behavior of the bandgaps of CuI and CuBr differs strongly from each other~\cite{Tanaka.2001}. While the bandgap of CuI decreases with increasing temperature, the reverse is true for CuBr, while for intermediate Br-contents of around 20\% the bandgap almost does not depend on the temperature. Thus, it is intuitively clear that the temperature affects the bowing behavior. Therefore, for a better comparability of our experimental results at room temperature with the DFT computations which assume zero temperature we extrapolate the obtained $E_0$ transition energies to low temperatures (T=\SI{10}{K}) using the temperature-dependence of the excitonic resonance for different alloy compositions published by Tanaka et al.~\cite{Tanaka.2001} (see Sec. IV in supplementary material for detailed discussion). Finally, taking into account both the excitonic and temperature effects, the bandgap bowing parameter at low temperatures is estimated to be \SI{0,49}{eV}. 
		In Sec.~VI of the supplementary material the influence of higher-energy configurations of \ce{CuBr_xI_{1-x}} are shown together with the averaged value of the bowing. Therefore, we can conclude that disorder plays a certain role for \ce{CuBr_xI_{1-x}}.



		To discuss further the origin of the observed bowing regarding the bandgaps, we follow the formalism introduced by Bernard and Zunger~\cite{Bernard_1987}. There the contributions to the experimentally observed bowing $b_\mathrm{exp}$ are split into a parameter that covers the effects, which are already included in an ordered alloy $b_\mathrm{I}$, and the ones which are introduced due to disorder effects $b_\mathrm{II}$:
		
		\begin{align}
			b_\mathrm{exp} &= b_\mathrm{I} + b_\mathrm{II} 
		\end{align} 
	
		Furthermore, $b_\mathrm{I}$ can be separated into three physical/chemical contributions. At first there is $b_\mathrm{VD}$, which covers effects due to volume deformations. Therefore, one compares energy levels $\epsilon$ for the perfect binaries with binaries deformed to the volume of the relaxed CuBr$_{0.5}$I$_{0.5}$ alloy (denoted here with the lattice parameter $a_{0.5}$):
		
		\begin{align}
			b_\mathrm{VD} &= 2\left[ \epsilon_\mathrm{CuI}(a_\mathrm{CuI}) + \epsilon_\mathrm{CuBr}(a_\mathrm{CuBr}) \right] \\ &\ \ \ \ - 2\left[ \nonumber  \epsilon_{\ce{CuI}}(a_{0.5}) + \epsilon_{\ce{CuBr}}(a_{0.5}) \right].\nonumber
		\end{align}
	
		Secondly, there is a parameter reflecting the effect, due to different chemical electronegativities $b_\mathrm{CE}$. Here the deformed binaries are compared with the 50\%:50\% alloy, where the internal atomic lengths $u$ are not relaxed (i.e. all atoms are in the positions which they would occupy in a perfect, undeformed zincblende crystal).
		
		\begin{align}
			b_\mathrm{CE} &= 2\left[ \epsilon_\mathrm{CuI}(a_{0.5}) + \epsilon_\mathrm{CuBr}(a_{0.5}) \right]\nonumber \\ &\ \ \ \  - 4 \epsilon_{\ce{CuBr_{0.5}I_{0.5}}}(a_{0.5}, u_{unrel}) .
		\end{align}
	
		At last, there is a parameter $b_\mathrm{S}$ covering the  effects arising when the system is allowed to fully relax the position of the atoms. So in this step the fully relaxed cell is compared to the cell where the atoms would sit in a perfect fcc cell:
		
		\begin{align}
			b_\mathrm{S} &= 4 \epsilon_{\ce{CuBr_{0.5}I_{0.5}}}(a_{0.5}, u_\mathrm{unrel}) - 4 \nonumber \epsilon_{\ce{CuBr_{0.5}I_{0.5}}}(a_{0.5}, u_\mathrm{rel})
		\end{align}
	
		In the end all these parameters simply add up to the entire bowing linked to ordered effects:
		
		\begin{align}
			b_\mathrm{I} &= b_\mathrm{VD} + b_\mathrm{CE} + b_\mathrm{S}
		\end{align}	
	
		The corresponding energy levels as well as the bowing parameters deduced by our calculations are given in Table~\ref{Table_bowingParameters}.

		\begin{table}[h]
			\centering
  			\caption{ Overview of the obtained values for the bandgap E\textsubscript{g} (eV) and split-off at the VBM $\Delta_0$ due to SOC for different configurations of the \ce{CuBr_xI_{1-x}} systems together with the calculated bowing parameters. The values for the alloys are for the lowest energy configuration. }
  			\label{Table_bowingParameters}
  			\begin{tabular}{@{\extracolsep{\fill}}lll}
    			\hline
    			Composition & $\mathrm{E_g} (\mathrm{eV})$ & $\Delta_0$ (eV) \\
				\hline
				$\epsilon_\mathrm{CuI}(a_\mathrm{CuI})$ & 3.13 & 0.68 \\
  				$\epsilon_\mathrm{CuBr}(a_\mathrm{CuBr})$ & 2.81 & 0.17 \\
   				$\epsilon_\mathrm{CuI}(a_{0.5})$  & 3.33 & 0.72 \\
   				$\epsilon_\mathrm{CuBr}(a_{0.5})$ & 2.71 & 0.14\\
   				$\epsilon_{\ce{CuBr_{0.5}I_{0.5}}}(a_{0.5}, u_\mathrm{unrel})$ & 2.91 & 0.46 \\
   				$\epsilon_{\ce{CuBr_{0.5}I_{0.5}}}(a_{0.5}, u_\mathrm{rel})$ & 2.83 & 0.46\\
  				\hline
   				$b_\mathrm{VD}$  & -0.20 & -0.02 \\
   				$b_\mathrm{CE}$  & 0.44 & -0.12 \\ 
   				$b_\mathrm{S}$   & 0.32 & 0.00 \\
  				\hline
   				$b_\mathrm{I}$  & 0.56 & -0.14 \\
  				\hline
  			\end{tabular}
		\end{table}

	\begin{figure*}[!htbp]
	\includegraphics[width=\textwidth]{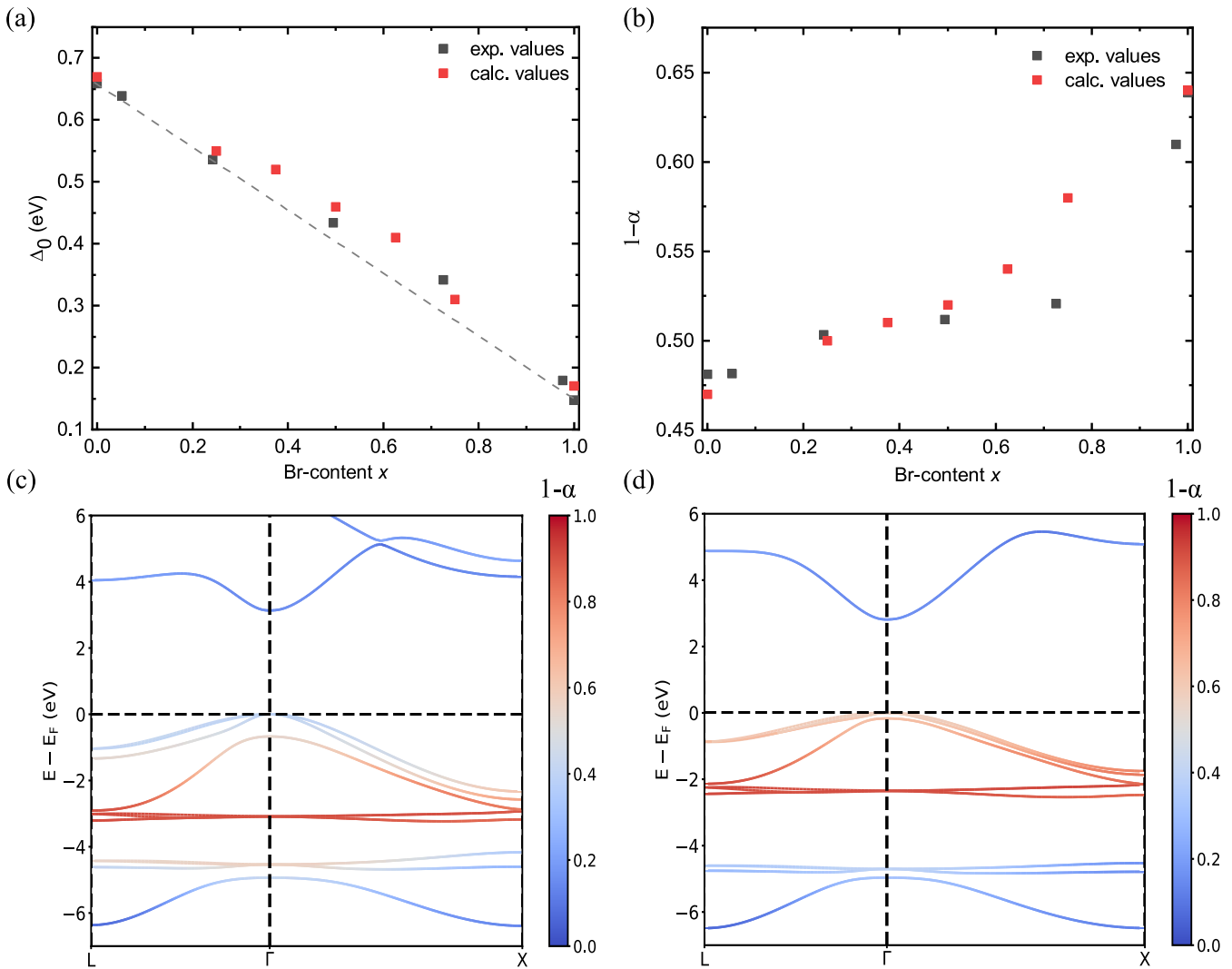} 
	\caption{ (a) Experimentally observed (black symbols) as well as theoretically calculated values (red symbols) for the spin-orbit splitting at the \textGamma-point $\Delta_0$ as function of the alloy composition. The dashed gray line indicates the almost linear decrease of the spin-orbit splitting with increasing Br-content. (b) The effect of Br-content on the contribution of copper d-orbitals to the VBM at the \textGamma-point. Again, the experimental data are indicated by black symbols, while the values obtained from the DFT calculations are shown in red. Fat band structures of CuI (c) and CuBr (d) highlighting the contribution of copper to the corresponding bands. The figures were made using pyprocar~\cite{Pyprocar}.}
	\label{fig4}
\end{figure*}

		We can see for the results regarding the bowing of the bandgap that the contribution of the volume deformation has a different sign than the ones due to the different chemical electronegativities and the structural effect. Although all three individual contributions are similar in their absolute magnitudes, the positive contributions $b_\text{CE}$ and $b_\text{S}$ predominate, resulting in an overall positive sign of the bowing parameter $b_\mathrm{I}$.  The differences with the previously mentioned results are that in the previous adjustment all calculated alloy compositions were considered, while here only the binary compounds and the alloy with the Br:I ratio of 1:1 were included.

	\subsection{Spin-orbit splitting $\Delta_0$}
	\label{sec:SOC}

		As next point we want to discuss the influence of the alloy composition on the value $\Delta_0$ of the spin-orbit splitting at the \textGamma-point. Experimentally we determined this value  as the energy difference of the peak positions E$_0$ and E$_0+\Delta_0$, in the simulations as direct energy difference calculated for the bands at the \textGamma-point (see Fig.~\ref{fig4} (a)). We expect an influence of the excitonic binding energy when we want to compare experiment with DFT. Information on the excitonic binding energies are given in Sec.~V of the supplementary material. The observed $\Delta_0$ values decrease almost linearly with increasing Br-concentration, as expected due to smaller one-electron spin-orbit splitting of bromine compared to iodine. We note that our DFT calculations provide a very good quantitative description of the spin-orbit coupling over the entire alloy composition. The reason that the values obtained for the alloy \ce{CuBr_xI_{1-x}} are significantly lower than those that can be expected for the zincblende structure based on the one-electron values for Br and I~\cite{Braunstein.1959} has been attributed in the past to the strong p-d hybridization of the valence bands~\cite{Cardona_1963}, with the d-orbitals making a negative contribution to the SOC~\cite{Shindo.1965}. In this context, a simple empirical model which links the different contributions of the copper and the halogen atoms to the wave functions at the \textGamma-point \ with the resulting strength of the SOC was suggested by Cardona~\cite{Cardona_1963}:
		
		\begin{align}
			\Delta_0 &= \frac{3}{2} \left( \alpha\Delta_\mathrm{halogen} - (1-\alpha) \Delta_\mathrm{metal}  \right)
			\label{eq:alpha}
		\end{align}

		Now with our calculations we can verify the validity of this formula. We estimated $\alpha$ experimentally by simply using the measured results for the corresponding values of $\Delta_0$ and calculate it directly using Eq.~\ref{eq:alpha} and the one-electron spin-orbit splitting parameters of \SI{0.1}{eV}, \SI{0.45}{eV} and \SI{0.94}{eV} for Cu, Br and I, respectively~\cite{Cardona_1963}. On the other hand we calculated also the contributions of both copper and the halogen atoms to the wave functions within DFT. The results for the obtained values of ($1-\alpha$) representing the contribution of the Cu-atoms to the VBM splitting are shown in Fig~\ref{fig4} (b). Here, again, good agreement can be observed between theory and experiment. It is therefore evident that the influence of copper d-states on the VBM increases with increasing Br-content. Thus, increasing p-d hybridization leads to an almost identical bandgap of CuBr and CuI, although a larger bandgap is expected for CuBr due to the larger electron affinity of bromine compared to iodine. Regarding the calculation of the contribution of the corresponding elements at the VBM we like to point out the following: In Fig~\ref{fig4} (c) and (d) we see the contribution of Cu d-states in the band structure as a fat band plot. This value is increasing for the uppermost valence bands when going to lower energies starting at the VBM. This means, that for extracting these values the inclusion of SOC is actually very important. \\

		Although, as mentioned above, the overall shape of the $\Delta_0$ curve of the \ce{CuBr_xI_{1-x}} alloy is very close to a linear function (see Fig.~\ref{fig4} (a)), a slight bowing behavior can also be observed here but with a different sign compared to the E\textsubscript{0} bowing. To analyze this further we repeated the calculation of the corresponding bowing parameters as it was done for the bowing of the bandgap (see Tab.~\ref{Table_bowingParameters}). We note that the bowing in this case is dominated mainly by the chemical contribution, which now has a negative sign, in contrast to the calculations performed for the bandgap bowing. The other contributions can be almost neglected here. Thus, we obtain a value for the $\Delta_0$ bowing parameter of \mbox{$b$=\SI{-0.14}{eV}} in reasonable agreement with experimental data.

	\section{Summary}
	
		In conclusion, we have shown that the optical properties of \ce{CuBr_xI_{1-x}} alloy can be systematically manipulated by changing the chemical alloy composition. The dielectric function of \ce{CuBr_xI_{1-x}} alloy thin films was determined by means of spectroscopic ellipsometry at room tempeature in the spectral range from \SI{0.7}{eV} to \SI{6.4}{eV}. First-principles band structure calculations were performed for ordered alloys and used for assignment of the experimentally observed spectral features to electronic transitions in different regions of the Brillouin zone involving restricted groups of bands. In contrast to the previously discussed assignment of the higher energy transitions~\cite{Cardona_1963}, we were able to show that the $\text{E}'_\text{0}$ transition in the mixed CuI-CuBr system does not occur at the \textGamma-point but near the L-point. The experimentally observed shift with composition of the E\textsubscript{0} transition at room temperature can be described by a bowing parameter \mbox{$b_{\text{exp}}\approx\SI{0.50}{eV}$}. The corresponding bowing behavior of the bandgap at low temperatures was estimated to a value of \SI{0.49}{eV} considering the exciton binding energy, as well as the temperature dependence of the bandgap as function of alloy composition. The DFT calculations of the lowest energy configuration is $b_\text{I} \approx$ \SI{0.56}{eV}. The ensemble average of all ordered configuration is \mbox{$b_\text{I} \approx \SI{1.00}{eV}$}. We expect the full disordered picture to be between those 2 values and therefor in good agreement with the experimental results. Furthermore, the bowing parameter was described in terms of different physical and chemical contributions: we show that effects due to different atomic electronegativities and structural contributions dominate in the investigated alloys. The spin-orbit splitting $\Delta_0$ was found to decrease from a value of \SI{660}{meV} for CuI to \SI{150}{meV} for CuBr. Based on our calculations, we find a significant contribution of Cu $d$-orbitals to the VBM, and this $p$-$d$ hybridization actually increases with increasing Br-content. Although the observed decrease in $\Delta_0$ is nearly linear, we find that the effect due to the different electronegativities of Br and I leads to a contribution to the bowing of the spin-orbit splitting with a different sign compared to the main bandgap contribution.

	\section{Acknowledgments}
		We gratefully acknowledge funding from the Deutsche Forschungsgemeinschaft (DFG, German Research Foundation) through FOR 2857 (Projects P02, P04, P05, and P06) - 403159832. E.K and M.S.B. acknowledge the Leipzig School of Natural Sciences BuildMoNa. M.S. and S.B. acknowledge the Leibniz Supercomputing Centre for providing computational resources (project pn68le).

	\section{References}
	\bibliographystyle{unsrt}

\end{document}